\documentclass[conference]{IEEEtran}
\IEEEoverridecommandlockouts

\usepackage{cite}
\usepackage{amsmath,amssymb,amsfonts}
\usepackage{graphicx}
\usepackage{textcomp}
\usepackage{xcolor}
\usepackage{booktabs}
\usepackage{array}
\usepackage{microtype}
\usepackage{url}
\usepackage{algorithm}
\usepackage{algorithmic}
\usepackage{comment}
\usepackage[absolute,overlay]{textpos}

\newcommand{\snp}{SN\,P}
\newcommand{\R}{\mathbb{R}}
\newcommand{\E}{\mathbb{E}}

\begin{document}
\renewcommand\IEEEkeywordsname{Keywords}
\setlength{\textfloatsep}{3pt}
\setlength{\abovecaptionskip}{3pt}
\setlength{\belowcaptionskip}{3pt}
\setlength{\abovedisplayskip}{3pt}
\setlength{\belowdisplayskip}{3pt}
\allowdisplaybreaks




\title{From Wireless SNNs to SN P Systems:\\A Low-Energy Rule-Based Conversion}

\author{%
  \IEEEauthorblockN{Pietro Savazzi\IEEEauthorrefmark{1}\IEEEauthorrefmark{2},
                    Mauro Marchese\IEEEauthorrefmark{1},
                    Anna Vizziello\IEEEauthorrefmark{1}\IEEEauthorrefmark{2},
                    Fabio Dell'Acqua\IEEEauthorrefmark{1}\IEEEauthorrefmark{2}}
  \IEEEauthorblockA{\IEEEauthorrefmark{1}
    Department of Electrical, Computer and Biomedical Engineering,
    University of Pavia, 27100 Pavia, Italy
\IEEEauthorblockA{\IEEEauthorrefmark{2}
    CNIT Consorzio Nazionale Interuniversitario per le Telecomunicazioni,
    Unità di Pavia, Italy}
    Email: \{name.surname\}@unipv.it, mauro.marchese01@universitadipavia.it }
}

\IEEEoverridecommandlockouts
\IEEEpubid{\makebox[\columnwidth]{979-8-3195-0489-0/26/\$31.00~\copyright2026 IEEE \hfill} 
\hspace{\columnsep}\makebox[\columnwidth]{ }}
\maketitle
\IEEEpubidadjcol
\begin{textblock*}{3cm}(17cm,1cm)
\fontsize{10}{12}\selectfont (Special Session)
\end{textblock*}

\begin{abstract}
Distributed wireless spiking neural networks (DWSNNs) are a
promising paradigm for energy-efficient edge inference in
resource-constrained environments such as wireless sensor networks
(WSNs). Yet, two limitations persist: their
internal decision process is opaque, and their residual energy
footprint remains a limiting factor for ultra-low-power deployments.
This paper proposes a systematic methodology to convert a trained
DWSNN into an equivalent Spiking Neural P (\snp{}) system, a biologically-inspired,
rule-based computational model drawn from membrane computing, by
extracting symbolic firing rules from the hidden-layer spike
activity. The resulting \snp{} system provides direct,
human-readable decision explanations while consuming three orders
of magnitude less energy than its parent SNN. Experiments on the
Neuromorphic MNIST (N-MNIST) dataset with a two-layer fully
connected SNN using phase encoding and Leaky Integrate-and-Fire
(LIF) neurons show that the \snp{} system retains approximately
84\% of the original classification accuracy
(73.77\% vs.\ 87.68\%) while the output layer connectivity decreases from 1000 to 120 class-specific connections. This complexity reduction is governed by a parameter related to the number of relevant hidden neurons per class that can be chosen according to a trade-off between computational complexity reduction and output accuracy. These results position \snp{} systems as
lightweight, interpretable surrogates for trained distributed wireless SNNs in
neuromorphic edge deployments.
\end{abstract}


\section{Introduction}
\label{sec:intro}

The proliferation of edge intelligence in wireless sensor networks
(WSNs), Internet of Things (IoT), and space communication
systems imposes stringent energy and latency budgets on individual
sensor nodes~\cite{Furano2020,Gamba2014}. Traditional deep neural
networks (DNNs), relying on dense floating-point matrix
operations, are ill-suited to such constraints. Spiking neural
networks (SNNs)~\cite{Neftci2019,Burkitt2006}, which communicate
through sparse, asynchronous spike events, offer a
biologically-motivated alternative that aligns naturally with
event-driven neuromorphic hardware.

Building on this foundation, distributed wireless SNNs
(DWSNNs)~\cite{Liu2024,Borsos2022} partition inference across
spatially distributed sensor nodes interconnected via spikes that could be efficiently implemented as impulse
radio (IR) links. Each node encodes local sensory data into spike
trains and transmits them to downstream nodes for collaborative
classification. Recent works~\cite{Savazzi2024}\cite{Savazzi2025} established
information-theoretic performance bounds for four neuromorphic
coding schemes: rate coding, time-to-first-spike (TTFS), phase
coding, and burst coding, showing that the choice of input
encoder significantly affects transmission efficiency, inference
accuracy, and energy consumption under additive white Gaussian
noise (AWGN) conditions.

Despite their efficiency, DWSNNs retain two fundamental
shortcomings. First, the inference logic is distributed across
thousands of continuously valued synaptic weights and membrane
potentials, rendering the model a \emph{black box}: no direct
human-interpretable rationale is available for individual
predictions. This opacity is increasingly unacceptable in
safety-critical and regulatory-compliant IoT applications.
Second, while spike-based communication is inherently sparser
than conventional DNN traffic, the total number of spikes
generated during inference is still non-trivial for
battery-operated or energy-harvesting sensor platforms, which must adhere to nanojoule-level energy budgets~\cite{Kucik2021}.

Membrane computing~\cite{Paun2000}, introduced by P\u{a}un,
offers a formal framework for distributed parallel computation
inspired by the chemistry of biological cells. Within this
framework, \emph{Spiking Neural P (\snp{}) systems}~\cite{Ionescu2006}
define neural-like computing devices whose elementary operations
are symbolically specified firing rules over discrete spike counts.
Unlike connectionist SNNs, whose behavior emerges from large-scale
numerical dynamics, \snp{} systems operate through formal rule
applications that are directly interpretable and formally
verifiable. Their event-driven, symbol-sparse computation is
naturally compatible with ultra-low-energy embedded platforms.

The main contribution of this paper is a systematic
data-driven methodology for converting a trained DWSNN into an
equivalent \snp{} system through rule extraction from hidden-layer
spike activity. The proposed conversion:
(i) preserves the dominant classification behavior of the original
    SNN with accuracy loss not exceeding 14 percentage points;
(ii) reduces inference energy by three orders of magnitude; and
(iii) yields a compact, symbolic model of 10 firing rules that
     explicitly identifies which neuronal pathways drive each
     class decision.

The remainder of this paper is organized as follows.
Section~\ref{sec:system} describes the DWSNN system model;
Section~\ref{sec:snn} details the SNN architecture and training, while 
section~\ref{sec:snp} introduces the \snp{} system formalism.
Section~\ref{sec:conversion} presents the conversion methodology
and energy analysis, and section~\ref{sec:results} reports simulation
results and discussion. Section~\ref{sec:conclusion} concludes.

\section{System Model}
\label{sec:system}

\begin{figure*}[t]
  \centering
  \includegraphics[width=6in]{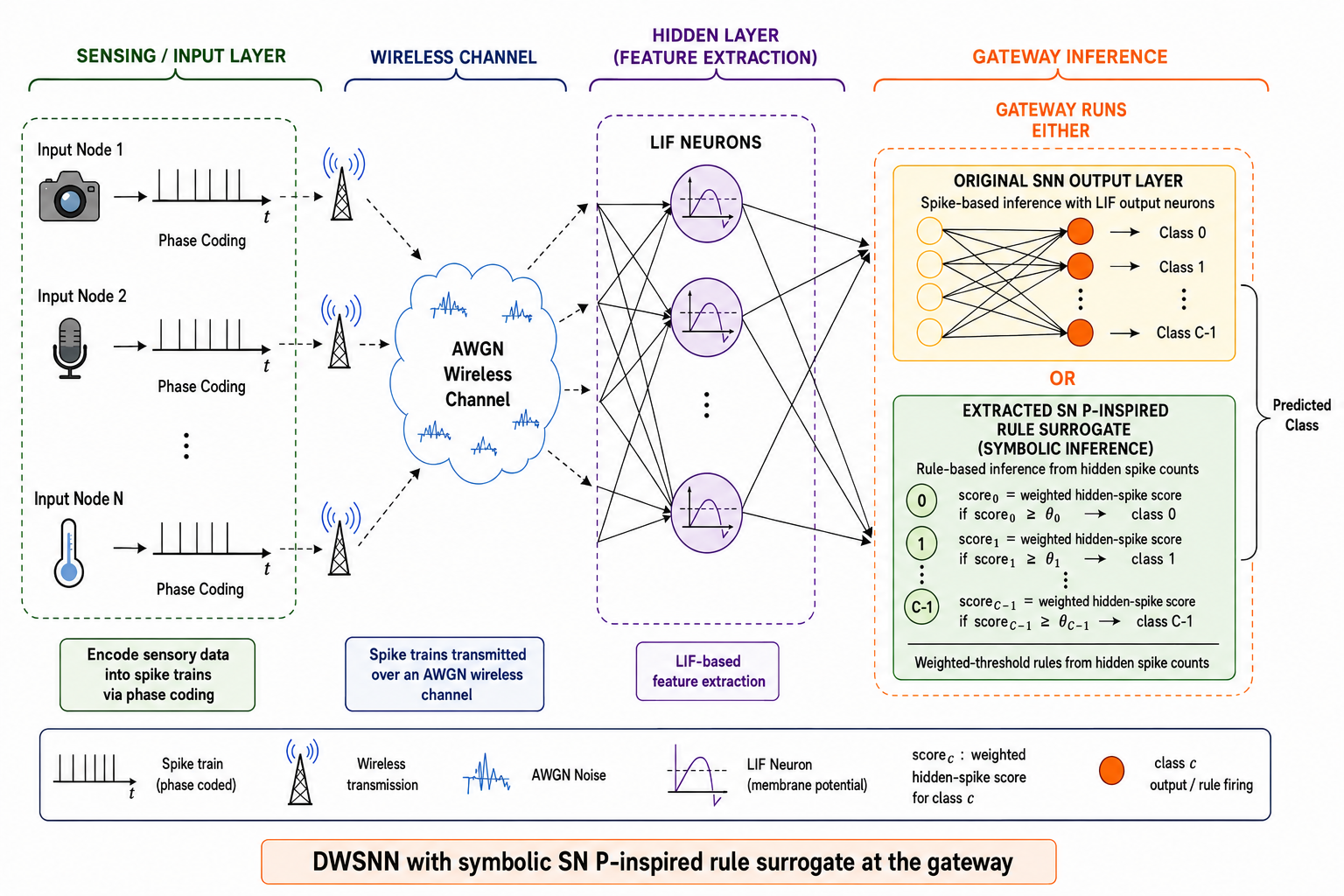}
  \caption{System model of the DWSNN with symbolic \snp{}
           inference at the gateway. Input nodes encode sensory
           data into spike trains via phase coding and transmit
           over an AWGN wireless channel. Hidden nodes perform
           LIF-based feature extraction. The gateway runs either
           the original SNN or the extracted \snp{} system.}
  \label{fig:system}
\end{figure*}

The system model, illustrated in Fig.~\ref{fig:system}, follows
the DWSNN architecture of~\cite{Savazzi2025}, extended to
incorporate a symbolic inference stage at the gateway.

\subsection{Functional Architecture}
The system comprises three functional levels:
\begin{itemize}
  \item \textbf{Input (sensing) nodes}: each node acquires a local
        measurement, normalizes it to $[0,1]$, and encodes it
        into a spike train $\mathbf{x}(t)\in\{0,1\}^{N_\text{in}}$
        using the selected neuromorphic coding scheme.
  \item \textbf{Hidden nodes}: these nodes receive encoded spike trains via
        the wireless channel and perform LIF-based processing to
        extract intermediate features.
  \item \textbf{Gateway (inference) node}: it aggregates hidden-node
        spike trains and performs classification, either via the
        full SNN output layer or the extracted \snp{} system.
\end{itemize}

\subsection{Wireless Channel Model}
All wireless links are modeled as AWGN channels. The signal
received at a downstream node is
\begin{equation}
  \mathbf{y}(t) = \;\mathbf{x}(t) + \mathbf{n}(t),
  \label{eq:channel}
\end{equation}
where $\mathbf{x}(t)$ is the transmitted spike vector,
$\mathbf{n}(t)\!\sim\!\mathcal{N}(\mathbf{0},\sigma^2\mathbf{I})$ is
independent Gaussian noise, and $\sigma^2$ is the
noise power. Following~\cite{Savazzi2025}, we
adopt \emph{phase coding} at the input layer: the $N_b$-bit
binary representation of each quantized amplitude generates
$N_b$ equally-spaced spikes, each weighted by its positional
bit value.

\subsection{Energy Model}
Following the IR transmission model of~\cite{Savazzi2025}, energy
is estimated from the total spike count $N_\text{spikes}$ as 
\begin{equation}
  \mathcal{E} = N_\text{spikes}\cdot\varepsilon_0,
  \label{eq:energy}
\end{equation}
with $\varepsilon_0 = 10^{-9}$~J per spike. This hardware-agnostic
baseline enables a fair comparison between architectures with
different computational strategies.

\section{Spiking Neural Networks}
\label{sec:snn}

\subsection{Neuron Model}\label{sec:neuronmodel}
Each neuron is modeled as a Leaky Integrate-and-Fire (LIF)
unit~\cite{Burkitt2006}. The membrane potential $u_i(t)$ evolves as
\begin{equation}
  u_i(t) = \beta\,u_i(t-1)\bigl(1 - z_i(t-1)\bigr) + h_i(t),
  \label{eq:lif}
\end{equation}
where $\beta\in(0,1)$ is the leak decay constant and $h_i(t)$
is the weighted synaptic input. The output spike is generated by
the Heaviside threshold operation
\begin{equation}
  z_i(t) = \Theta\bigl(u_i(t) - \vartheta\bigr),
  \label{eq:thresh}
\end{equation}
with $\vartheta$ the firing threshold. Firing resets the membrane
potential through the factor $(1-z_i(t-1))$ in~\eqref{eq:lif},
implementing a standard soft-reset mechanism.

\subsection{Network Architecture}
We adopt a two-layer fully connected SNN with $N_h=100$ hidden LIF
neurons and $C=10$ output neurons. Denoting
$\mathbf{W}_1\!\in\!\R^{N_h\times N_\text{in}}$ and
$\mathbf{W}_2\!\in\!\R^{C\times N_h}$ as the weight matrices,
the forward pass at time step $t$ is
\begin{align}
  \mathbf{h}_1(t) &= \mathbf{W}_1\,\mathbf{y}(t) + \mathbf{n}_1(t),
                    \label{eq:h1}\\
  \mathbf{h}_2(t) &= \mathbf{W}_2\,\mathbf{z}_1(t) + \mathbf{n}_2(t),
                    \label{eq:h2}
\end{align}
where $\mathbf{z}_1(t)$ are the hidden-layer spikes (see Section \ref{sec:neuronmodel}) and
$\mathbf{n}_k(t)\!\sim\!\mathcal{N}(\mathbf{0},\sigma^2\mathbf{I})$
models AWGN at each layer. Both $\mathbf{h}_1(t)$ and $\mathbf{h}_2(t)$ are then fed
through the LIF dynamics of~\eqref{eq:lif}--\eqref{eq:thresh}.

\subsection{Training}
Training uses surrogate gradient descent via backpropagation
through time (BPTT)~\cite{Neftci2019}, approximating the
non-differentiable step $\Theta(\cdot)$ with a smooth sigmoid.
The composite loss combines rate-coded cross-entropy and a
mean-square reconstruction term
\begin{equation}
  \mathcal{L} = \mathcal{L}_{\mathrm{CE}}(\mathbf{z}_2)
              + \mathcal{L}_{\mathrm{recon}}\!\left(\mathbf{x},
                \hat{\mathbf{x}}\right),
  \label{eq:loss}
\end{equation}
where
$\hat{\mathbf{x}} = \frac{1}{T}\sum_{t=1}^{T}\mathbf{W}_r\,\mathbf{z}_1(t)$
is the reconstructed input image obtained through a linear decoder
$\mathbf{W}_r\!\in\!\R^{N_\text{in}\times N_h}$. The Adam
optimizer~\cite{Kingma2017} is used with learning rate $10^{-3}$
and batch size 64.

\section{Spiking Neural P Systems}
\label{sec:snp}

\subsection{Formal Definition}
Spiking Neural P (\snp{}) systems~\cite{Ionescu2006} are a class
of parallel distributed computing devices, introduced within the
membrane computing framework of P\u{a}un~\cite{Paun2000}, in which
the fundamental computational units are neurons communicating
exclusively through spike events. An \snp{} system of degree
$m\geq 1$ is defined as
\begin{equation}
  \Pi = \bigl(O,\;\sigma_1,\ldots,\sigma_m,\;
             \mathrm{syn},\;\mathrm{in},\;\mathrm{out}\bigr),
  \label{eq:snp_def}
\end{equation}
where $O=\{a\}$ is the singleton spike alphabet; each neuron
$\sigma_i=(q_i,R_i)$ holds an initial spike count $q_i\geq 0$
and a finite set of firing rules $R_i$; $\mathrm{syn}\subseteq
\{1,\ldots,m\}^2$ is the directed synapse set; and
$\mathrm{in},\,\mathrm{out}$ designate the input/output neurons.

Each rule $r\in R_i$ has the canonical form
\begin{equation}
  E \;/\; a^c \;\rightarrow\; a^p\,;\;d,
  \label{eq:rule}
\end{equation}
where $E$ is a regular expression over $\{a\}$ specifying the
applicable spike configuration, $c\geq 1$ spikes are consumed,
$p\geq 0$ spikes are produced, and $d\geq 0$ is the synaptic delay.
Rule $r$ is \emph{applicable} when the neuron's current spike count
matches $E$ and is at least $c$.

\subsection{Properties Relevant to Edge Inference}
Compared with connectionist SNNs, \snp{} systems present three
properties particularly attractive for energy-constrained wireless
deployments:

\begin{enumerate}
  \item \textbf{Symbolic interpretability:} Each rule $r\in R_i$
        is a human-readable, formally verifiable condition: it
        specifies exactly which spike pattern triggers class
        assignment, enabling real-time auditability.
  \item \textbf{Extreme sparsity:} Classification is performed
        by evaluating $C$ rules in parallel; only the matching
        rule fires. The average number of rule evaluations per
        sample is markedly lower than the spike count of the
        parent SNN.
  \item \textbf{Biological plausibility:} Like SNNs, \snp{}
        systems are event-driven and operate on binary spike
        signals, maintaining compatibility with neuromorphic
        hardware such as Intel Loihi or IBM TrueNorth.
\end{enumerate}

\section{SNN-to-\snp{} Conversion}
\label{sec:conversion}

\subsection{Rule Extraction Algorithm}
\begin{figure*}[t]
  \centering
  \includegraphics[width=6.5in]{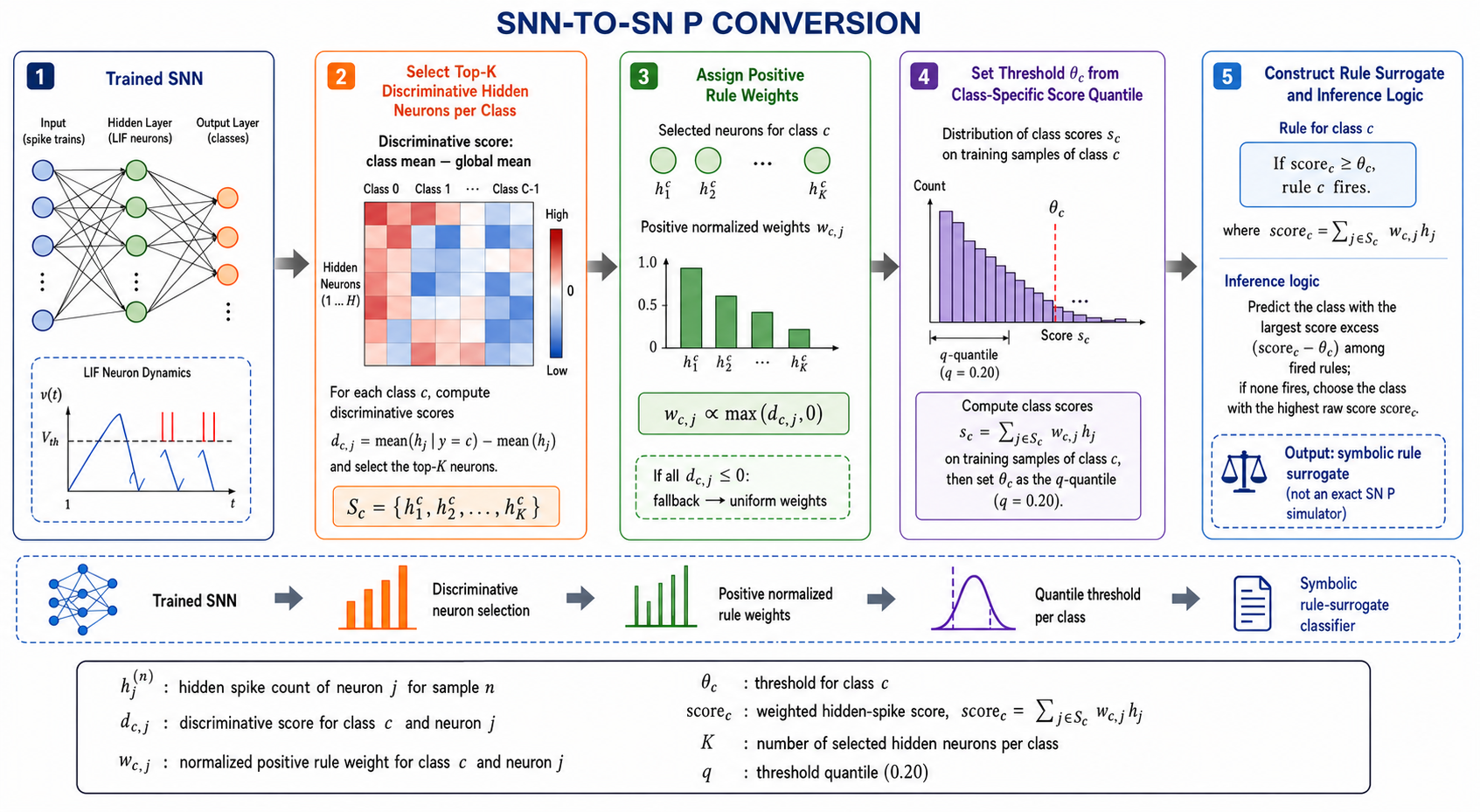}
  \caption{Block diagram of the data-driven SNN-to-\snp{} conversion. (1) Hidden spike counts are gathered from the trained SNN. (2) The top-$K$ discriminative hidden neurons per class are selected using the difference between class-specific and global activation means. (3) Positive rule weights are computed and normalized. (4) Firing thresholds $\theta_c$ are estimated via a class-specific score $q$-quantile ($q=0.20$). (5) The final symbolic rule surrogate classifies inputs using a two-stage maximum excess and raw score fallback inference logic.}
  \label{fig:conversion}
\end{figure*}

The proposed conversion extracts a symbolic, rule-based surrogate classifier from the hidden spike patterns generated by the trained SNN. Instead of relying directly on the network's connectionist weights, the algorithm identifies class-specific discriminative hidden neurons, normalizes their activation contribution, and defines a robust firing threshold using score quantiles. The complete pipeline is formalized in Algorithm~\ref{alg:extract} and illustrated in Fig.~\ref{fig:conversion}.

\begin{algorithm}[t]
\caption{SNN-to-\snp{} Rule Extraction}
\label{alg:extract}
\begin{algorithmic}[1]
\REQUIRE Training hidden spike counts $\{h_j^{(n)}\}$ for sample $n$ and neuron $j$; ground-truth labels $\{y^{(n)}\}$; hyperparameter $K$; threshold quantile $q$.
\ENSURE  Set of symbolic rules and parameters $\{(\mathcal{S}_c, \mathbf{w}_c, \theta_c)\}_{c=0}^{C-1}$
\FOR{$c = 0$ \TO $C-1$}
  \STATE Compute discriminative scores $d_{c,j}$ via~\eqref{eq:disc_score} for all hidden neurons
  \STATE $\mathcal{S}_c \leftarrow \operatorname{argtop}_K\bigl(d_{c,\cdot}\bigr)$
  \STATE Assign and normalize positive rule weights $w_{c,j}$ via~\eqref{eq:weights_snp}
  \STATE Calculate class scores $s_c^{(n)} = \sum_{j\in\mathcal{S}_c} w_{c,j} h_j^{(n)}$ for all training samples where $y^{(n)} = c$
  \STATE Set $\theta_c \leftarrow \operatorname{quantile}\bigl(\{s_c^{(n)}\}, q\bigr)$
  \STATE Generate rule: $E_c \;/\; a^{\theta_c} \;\rightarrow\; a_c$
\ENDFOR
\RETURN $\{(\mathcal{S}_c, \mathbf{w}_c, \theta_c)\}_{c=0}^{C-1}$
\end{algorithmic}
\end{algorithm}

The formal conversion steps are detailed below:

\begin{enumerate}
    \item \textbf{Discriminative Neuron Selection:} For each class $c$, the relevance of the hidden layer units is evaluated by measuring how much a neuron's average spike activity during class $c$ presentation deviates from its global baseline activity. The discriminative score $d_{c,j}$ for neuron $j$ is defined as
    \begin{equation}
      d_{c,j} = \mathbb{E}[h_j \mid y = c] - \mathbb{E}[h_j],
      \label{eq:disc_score}
    \end{equation}
    where $h_j$ is the total spike count of hidden neuron $j$ over the observation window $T$. The support set $\mathcal{S}_c$ is formed by selecting the top-$K$ hidden neurons with the highest $d_{c,j}$.

    \item \textbf{Positive Weight Allocation:} To reflect excitatory \snp{} rule behaviors, only positive discriminative contributions are retained. The rule weights $w_{c,j}$ for $j \in \mathcal{S}_c$ are derived by rectifying and scaling the scores
    \begin{equation}
      w_{c,j} = \frac{\max(d_{c,j}, 0)}{\sum_{k \in \mathcal{S}_c} \max(d_{c,k}, 0)}.
      \label{eq:weights_snp}
    \end{equation}
    If $\max(d_{c,j}, 0) = 0$ for all selected neurons, the algorithm defaults to uniform weights ($w_{c,j} = 1/K$).

    \item \textbf{Quantile Threshold Estimation:} For each training sample belonging to class $c$, an aggregate activation score is computed as $s_c = \sum_{j \in \mathcal{S}_c} w_{c,j} h_j$. To ensure the rule activates reliably even under noisy input conditions, the symbolic firing threshold $\theta_c$ is established as the $q$-quantile of the within-class score distribution
    \begin{equation}
      \theta_c = \operatorname{quantile}\bigl(\{s_c \mid y = c\}, \, q\bigr),
      \label{eq:threshold_quantile}
    \end{equation}
    where $q \in (0, 1)$ is a tunable hyperparameter (configured to $0.20$ by default). 

    \item \textbf{Two-Stage Symbolic Inference:} At inference time, the system computes the scores $s_c$ for all $C$ classes in parallel. A rule is considered triggered (\textit{fired}) if $s_c \geq \theta_c$. The final classification decision $c^*$ follows a two-stage selection policy
    \begin{equation}
      c^* = \begin{cases} 
        \displaystyle\arg\max_{c \in \mathcal{F}} \; (s_c - \theta_c) & \text{if } \mathcal{F} \neq \emptyset, \\ 
        \displaystyle\arg\max_{c} \; s_c & \text{if } \mathcal{F} = \emptyset,
      \end{cases}
      \label{eq:inference_logic}
    \end{equation}
    where $\mathcal{F} = \{c \mid s_c \geq \theta_c\}$ denotes the set of fired rules. If multiple rules fire, the system picks the class yielding the highest threshold excess; if no rule satisfies the firing condition, the network falls back to the maximum raw score.
\end{enumerate}

The structural complexity reduction at the output stage can be quantified independently of the energy model by comparing the number of class-specific connections used by the original SNN and by the extracted \snp{}-inspired rule surrogate. Recalling that $N_h$ denotes the total number of hidden neurons, $C$ the number of output classes, and $K$ the number of hidden neurons selected for each class-specific rule. For class $c$, the selected support set is denoted by $S_c$, with $\lvert S_c\rvert=K$. The fully connected SNN output layer contains
\begin{equation}
    N_{\mathrm{conn}}^{\mathrm{SNN}} = N_h C
\end{equation}
hidden-to-output connections, whereas the rule surrogate uses
\begin{equation}
    N_{\mathrm{conn}}^{\mathrm{SNP}} = KC
\end{equation}
class-specific connections. Therefore, the fraction of retained output-stage connectivity is
\begin{equation}
    \frac{N_{\mathrm{conn}}^{\mathrm{SNP}}}
         {N_{\mathrm{conn}}^{\mathrm{SNN}}}
    =
    \frac{KC}{N_hC}
    =
    \frac{K}{N_h},
    \label{eq:connectivity_ratio}
\end{equation}
and the corresponding connectivity reduction is
\begin{equation}
    \eta_{\mathrm{conn}}
    =
    1-
    \frac{N_{\mathrm{conn}}^{\mathrm{SNP}}}
         {N_{\mathrm{conn}}^{\mathrm{SNN}}}
    =
    1-\frac{K}{N_h}.
    \label{eq:connectivity_reduction}
\end{equation}
For the considered architecture, with $N_h=100$, $C=10$, and $K=12$, the number of output-stage connections decreases from $1000$ to $120$, corresponding to $\eta_{\mathrm{conn}}=0.88$, i.e., an $88\%$ reduction. Each hidden-to-rule connection is counted separately for every class, even when the same hidden neuron belongs to the support sets of multiple rules. This result quantifies the structural sparsification of the classifier and is independent of any assumption regarding the energy cost per spike or rule operation. A corresponding energy reduction cannot be inferred directly from \eqref{eq:connectivity_ratio}, since it also depends on the target hardware, memory-access costs, synaptic-event processing, threshold evaluations, output-rule firings, and static power consumption.

\section{Numerical Results}
\label{sec:results}

\subsection{Experimental Setup}

Experiments are conducted on the Neuromorphic MNIST (N-MNIST) dataset~\cite{Orchard2015}, a spiking adaptation of the standard handwritten digit benchmark captured via an asynchronous time-based image sensor (ATIS). The SNN architecture is implemented in snnTorch~\cite{Eshraghian2023}; the detailed network configuration, coding parameters, and training hyperparameters are summarized in Table~\ref{tab:snn_config}.

\subsection{Classification Performance and Energy Consumption}
Table~\ref{tab:results_unified} reports the test accuracy and inference energy on N-MNIST for both models. The SNN achieves $87.68\%$ accuracy, which is consistent with results for comparable architectures under the noise conditions considered in~\cite{Savazzi2025}. The extracted \snp{} system attains $73.77\%$ accuracy, retaining $\frac{73.77}{87.68}\approx 84.1\%$
of the original classification capability. Although not reported in the table, the symbolic model exhibits a prediction agreement of $77.66\%$ with the baseline SNN. This indicates that approximately four out of five SNN decisions are reproduced by the symbolic model, confirming that the extracted rules successfully capture the dominant inference mechanisms of the trained network.

The accuracy gap of approximately 14 percentage points reflects the deliberate compression of the \snp{} rule extraction: relying on $K=12$ out of 100 hidden neurons per class discards distributed fine-grained information encoded in the remaining synaptic weights. This is the quantifiable cost of achieving symbolic interpretability and computational sparsity.

Regarding computational efficiency, the inference energy is computed via~\eqref{eq:energy} over the entire N-MNIST test set. As shown in Table~\ref{tab:results_unified}, the \snp{} system achieves an energy reduction of approximately $50\times$ lower energy usage compared to the baseline SNN. This result is based on a rough estimation of the consumed energy, assigning 1 nJ to an SNN spike and 20 pJ to a heterogeneous \snp{} surrogate event. These decisions are based on information from available hardware platform implementations of neuromorphic spiking networks \cite{Davies2018}.

\begin{table}[t]
\centering
\caption{Simulation, network architecture, and training configuration details.}
\label{tab:snn_config}
\begin{tabular}{ll}
\toprule
\textbf{Parameter / Component} & \textbf{Value / Specification} \\
\midrule
Dataset & N-MNIST \cite{Orchard2015} \\
Input Frame Size & $34 \times 34 \times 2 = 2312$ elements \\
Software Framework & snnTorch \cite{Eshraghian2023} \\
\midrule
\textbf{Coding \& Timing} & \\
Coding Scheme & Phase coding \\
Bit Resolution ($N_b$) & 8 bits \\
Spreading Factor ($N_i$) & 64 \\
Time Steps ($T$) & 64 \\
\midrule
\textbf{Architecture} & \\
Network Structure & Fully connected $2312 \to 100 \to 10$ \\
Neuron Model & LIF ($\beta = 0.95$) \\
\midrule
\textbf{Training} & \\
Epochs & 5 \\
Optimizer & Adam ($\eta = 10^{-3}$, batch size 64) \\
Noise & AWGN ($\sigma^2 = 3.0$) \\
\midrule
\textbf{Extraction} & \\
Dominant Hidden Neurons ($K$) & 12 per class \\
Active Synapses & $C \times K = 120$ \\
Symbolic Firing Rules & 10 \\
\bottomrule
\end{tabular}
\end{table}

\begin{table}[t]
  \centering
  \caption{Classification Performance and Inference Energy on N-MNIST}
  \label{tab:results_unified}
  \begin{tabular}{lcc}
    \toprule
    \textbf{Model} & \textbf{Accuracy (\%)} & \textbf{Energy, $\mathcal{E}$ (J)} \\
    \midrule
    SNN            & 87.68                  & $3.16 \times 10^{-2}$ \\
    \snp{} System  & 73.77                  & $6.24 \times 10^{-4}$ \\
    \bottomrule
  \end{tabular}
\end{table}

\subsection{Symbolic Rule Analysis}

The \snp{} system consists of 10 firing rules, one per digit class.
Each rule is anchored to a support set of $K=12$ hidden neurons.
As a representative example, the rule for class~0 reads
\begin{equation}
  E_0 \;/\; a^{51.72} \;\rightarrow\; a_0,
\end{equation}
with dominant neuron support consisting of 
$\mathcal{S}_0 = \{50,47,11,0,76,19,72,62,45,30,70,84\}$.
This representation directly identifies which hidden neurons are
most informative for recognizing digit~``0,'' providing a
degree of transparency entirely absent in the original SNN.
In a distributed wireless deployment, such compact rules could
be transmitted alongside inference results to enable real-time
auditability of the gateway classifier, a property increasingly
demanded in safety-critical and regulatory-compliant IoT
applications.


The hidden layer, trained jointly via the reconstruction term
in~\eqref{eq:loss}, achieves an average input reconstruction
MSE of $3.62\times10^{-4}$ on the test set. The observed low error
confirms that the 100-neuron latent space retains a compact yet
information-dense representation of the N-MNIST event frames,
validating the quality of the activations used as the basis for
rule extraction.


Finally, Table~\ref{tab:tradeoff} summarizes the trade-off between the two
models across several system-level dimensions. The \snp{} system
sacrifices approximately 14 percentage points of accuracy while, on the other hand, attaining full symbolic explainability, a $\approx\!50\times$
energy reduction, and using a drastically simpler inference engine.

\begin{table}[t]
  \centering
  \caption{Trade-off Summary: SNN vs.\ \snp{} System}
  \label{tab:tradeoff}
  \begin{tabular}{lcc}
    \toprule
    \textbf{Property}        & \textbf{SNN}          & \textbf{\snp{} System}\\
    \midrule
    Interpretability         & Low                   & High (symbolic rules) \\
    Sparsity                 & Distributed           & Extremely sparse      \\
    Active connections       & $N_h\cdot C=1000$   & $K\cdot C=120$      \\
    Biological plausibility  & High                  & High                  \\
    Hardware complexity      & Moderate              & Very low              \\
    \bottomrule
  \end{tabular}
\end{table}

From the information-theoretic perspective of~\cite{Savazzi2025},
it is noteworthy that phase coding, identified as a competitive
scheme in terms of mutual information and inference accuracy,
provides a structured, deterministic spike representation
particularly amenable to rule extraction. The bit-weighted nature
of phase-coded spikes maps naturally to the symbolic threshold
conditions of \snp{} rules, suggesting an intrinsic compatibility
between the two paradigms that may be exploited in future joint
coding-and-rule-extraction design.

\section{Conclusion}
\label{sec:conclusion}

This paper has presented and experimentally validated a methodology
for converting a distributed wireless spiking neural network into
an equivalent Spiking Neural P (\snp{}) system via data-driven
rule extraction from hidden-layer spike activity. The conversion
achieves $73.77\%$ classification accuracy on N-MNIST
($84.1\%$ of the SNN performance), $77.66\%$ decision agreement
with the parent SNN, a $\approx\!50\times$ energy reduction, and full symbolic interpretability
through 10 compact firing rules. These results establish \snp{}
systems as viable, ultra-low-energy, and interpretable surrogates
for trained wireless SNNs in neuromorphic edge deployments.

Future work will explore multi-rule-per-class extraction,
temporal rule patterns exploiting spike timing, joint
optimization of neuromorphic coding and rule extraction
within the mutual-information framework of~\cite{Savazzi2025},
and hardware-aware rule compression targeting Intel Loihi
and analog neuromorphic platforms.

\bibliographystyle{IEEEtran}
\bibliography{references}

\end{document}